\documentclass[%
 reprint,
superscriptaddress,
 amsmath,amssymb,
 aps,
prb,
]{revtex4-2}

\usepackage{graphicx}
\usepackage{dcolumn}
\usepackage{bm}
\usepackage{mathtools}
\usepackage[bb=boondox]{mathalfa} 

\usepackage{amsmath}
\newcommand{\thickbar}[1]{\mathbf{\bar{\text{$#1$}}}}
\usepackage{mathrsfs}  
\usepackage{graphicx} 
\usepackage{caption}
\usepackage{float}
\usepackage{color}
\usepackage{xcolor}

\usepackage{booktabs} 
\usepackage{array} 
\usepackage{paralist} 
\usepackage{verbatim} 
\usepackage{subfig} 
\usepackage{float} 
\usepackage{hyperref}
\hypersetup{colorlinks=true,linkcolor=blue,urlcolor=blue}

\usepackage{xcolor}
\newif\ifproofread

\begin{document}

\preprint{APS/123-QED}

\title{{\color{black}  Two-mode Floquet Redfield quantum master approach for quantum transport}}

\author{Vahid Mosallanejad}
\email{vahid@westlake.edu.cn} 
\affiliation{Department of Chemistry, School of Science, Westlake University, Hangzhou, Zhejiang 310024, China}
\affiliation{Institute of Natural Sciences, Westlake Institute for Advanced Study, Hangzhou, Zhejiang 310024, China}

\author{Wenjie Dou}
\email{douwenjie@westlake.edu.cn} 
\affiliation{Department of Chemistry, School of Science, Westlake University, Hangzhou, Zhejiang 310024, China}
\affiliation{Institute of Natural Sciences, Westlake Institute for Advanced Study, Hangzhou, Zhejiang 310024, China}
\affiliation{Department of Physics, School of Science, Westlake University, Hangzhou, Zhejiang 310024, China}

\date{\today}
\begin{abstract}
Simultaneous driving by two periodic oscillations yields a practical technique for further engineering quantum systems. For quantum transport through mesoscopic systems driven by two strong periodic terms, a non-perturbative Floquet-based quantum master equation (QME) approach is developed using a set of dissipative time-dependent terms and the reduced density matrix of the system. This work extends our previous Floquet approach for transport through quantum dots (at finite temperature and arbitrary bias) driven periodically by a single frequency. In a pedagogical way, we derive explicit time-dependent dissipative terms. Our theory begins with the derivation of the two-mode Floquet Liouville-von Neumann equation. We then explain the second-order Wangsness-Bloch-Redfield QME with a slightly modified definition of the interaction picture. Subsequently, the two-mode Shirley time evolution formula is applied, allowing for the integration of reservoir dynamics. Consequently, the established formalism has a wide range of applications in open quantum systems driven by two modes in the weak coupling regime. The formalism's potential applications are demonstrated through various examples.

\end{abstract}
\maketitle
\section{\label{sec:1}Introduction}
Quantum transport through mesoscopic nanostructures exhibits a range of remarkable static characteristics, including quantized conductance~\cite{van1988quantized,wharam1988one, van1991quantum}, Coulomb blockade~\cite{fulton1987observation}, quantum interference~\cite{yacoby1991interference,scott1989conductance, bergmann1984weak}, Kondo effect~\cite{kondo1964resistance,goldhaber1998kondo}, and spin filtering~\cite{meservey1994spin}. 
Meanwhile, dynamic effects such as quantum coherence, polaron formation~\cite{ku2007quantum,mishchenko2015mobility,de1997dynamical}, molecular vibronic coupling~\cite{zobel2021surface}, etc., have attracted interest in the solid-state physics and quantum chemistry communities. 
Dynamical control, for example, by applying oscillating fields to the original systems, allows for the discovery of new properties unattainable by only altering static experimental conditions.
Very recently, there has been an increasing interest in controlling the dynamics of quantum systems or chemical reactions by driving the system with more than only one frequency, which may described as multi-mode Floquet engineering~\cite{yan2023multiphoton, barriga2024floquet,pan2017absorption,kubel2020probing,gustin2021high,koski2018floquet}. 
For instance, mediated by the strong spin-orbit interactions in a hole silicon spin qubits, the proposed phase-driving technique couples a secondary frequency (radio wave) to the qubit phase to reduce the susceptibility of the qubit to the noise~\cite{bosco2023phase}. To the best of our knowledge, there are no theoretical quantum transport reports (on open quantum systems) where the dot is driven with two frequencies. The interplay between two simultaneous drivings (specified by two amplitudes, two frequencies, and two initial phases) and the strength of electronic dissipation is intriguing from a theoretical standpoint. 
Several theoretical approaches have been developed to understand both the steady-state and the dynamical aspects associated with transport effects mentioned above. 
Among these are Landauer-B{\"u}ttiker~\cite{datta1997electronic,datta2005quantum}, generalized quantum master equation~\cite{cohen2011memory}, non-equilibrium Green's function~\cite{haug2008quantum}, and Hierarchal equation of motion (HEOM)~\cite{jin2008exact}. The HEOM is inherently time-dependent and capable of capturing non-Markovian effects, but it is computationally expensive~\cite{ye2016heom}. 
Generally speaking, no single theoretical quantum transport approach can capture all the transport features with the same level of simplicity. 
In previous work, we developed two types of Floquet-based quantum master equations (QMEs) to simulate transport properties in a periodically driven open quantum system, namely the Hilbert space QME and the Floquet space QME~\cite{mosallanejad2024floquet}.
It has been observed that the number of quantized current plateaus changes under the strong coupling regime when the driving amplitude is significantly larger than the energy difference between the two levels and the driving frequency is in resonance.
Floquet space QME has been used within the surface hopping algorithm to address the chemical processes of molecules under time-periodic driving near a metal surface~\cite{wang2024nonadiabatic}.
{\color{black} 
This paper is organized as follows. \autoref{sec:2} presents the methodology through several subsections. In the first and second subsections, we demonstrate the existence of a well-defined Floquet Liouville-von Neumann (LvN) equation and the relevant time-evolution operator for a Hamiltonian driven by two independent frequencies, whether commensurate or not. In the third and fourth subsections, we introduce the model Hamiltonian and the Redfield quantum master equation, which serves as our main transport framework. In the fifth to seventh subsections, we apply three key simplification steps to arrive at simplified dissipative terms, enabling simulation of the reduced density matrix dynamics. In the last subsection, we explain how to evaluate observables. In \autoref{sec:3}, we apply our new methods to a driven two-level system (Anderson model) weakly coupled to two thermal baths (left and right terminals), where the difference in electrochemical potentials between the baths controls the current. Time-dependent and quasi-steady-state current expectation values are calculated for several two-frequency driving scenarios. Finally, concluding remarks are provided in \autoref{sec:4}.
}
\section{\label{sec:2} formalism}
\subsection{ \label{sec2_subsec:0} Two-mode Floquet Liouville-von Neumann Equation: closed systems}
The Liouville von Neumann (LvN) for the density operator reads as: 
\begin{eqnarray}
\label{eq:1}
{\color{black} 
\frac{d\hat{\rho}(\bar{t})}{d\bar{t}} 
= \frac{-i}{\hbar}[\hat{H} (\bar{t}), \hat{\rho}(\bar{t})].
}
\end{eqnarray}
{\color{black}  We then scale the time as $\bar{t}/\hbar\!\rightarrow{t}$ (equivalent to $\hbar\!=\!1$)}. One can transform the original LvN equation into a Floquet counterpart, in the Floquet space, for any perfectly periodic single mode Hamiltonian, $\hat{H} (t)\!=\!\hat{H}(t+T)$, as
\begin{eqnarray}
\label{eq:2}
\frac{d \hat{\rho}^{F}(t)}{d t} = -{i} \left[\hat{H}^{F}, \hat{\rho}^{F}(t)\right].
\end{eqnarray}
{\color{black} The advantage of Floquet LvN is that it allows us to program the dynamics, using a time-independent Floquet Hamiltonian, $\hat{H}^{F}$.}
For the single-mode driven Hamiltonian, details on how to derive Eq.~(\ref{eq:2}) are given in Refs.~\cite{ivanov2021floquet,mosallanejad2023floquet}. 
{\color{black}  Multi-mode Floquet theory, however,  was initially introduced in an ad hoc manner~\cite{ho1983semiclassical}. Although it has been validated recently~\cite{poertner2020validity}, to the best of our knowledge, no rigorous derivation exists for the density-based two-mode Floquet theory. Focusing on $\hat{H}^{F}$ and recognizing the complexity of this transformation under a two-mode driven Hamiltonian, $\hat{H}(w_1,w_2,t)\!\equiv\!\hat{H}(t)$, we break the process into three steps:}
{\color{black}  (I) expanding the Hamiltonian and density operator by the 2D complex Fourier series,} 
(II) transformation of LvN into its Fourier representation by introducing four algebraic operators, and (III) transformation from the Fourier representation to the Floquet representation. {\color{black} We will also highlight the similarities and differences between the single-mode and two-mode Floquet versions.}

{\color{black}  Step (I) begins by expanding both the time-dependent Hamiltonian and density operators as }
\begin{eqnarray}
\label{eq:3}
\hat{H}(t)&=&\sum_{mn}^{} \hat{H}^{mn} 
\,
e^{i n \omega_1 t} 
e^{i m \omega_2  t}, 
\\
\label{eq:4}
\hat{\rho}(t)&=&\sum_{mn}^{} \hat{\rho}^{mn}(t) 
\,
e^{i n \omega_1  t}
e^{i m \omega_2  t}.
\end{eqnarray} 
{\color{black}  Coefficient operators $\hat{H}^{mn}$ determined by }
\begin{eqnarray}
\label{eq:5}
\begin{aligned}
\hat{H}^{mn}\!=\! \frac{1}{T_{1}T_{2}} \int_{0}^{T_2}\!\!\!\!dt_2\!\!\int_{0}^{T_1}\!\!\!\!dt_1 \hat{H}(t_1,t_2) e^{-in\omega_1 t_1} e^{-i m \omega_2 t_2},~~~~~~
\end{aligned}
\end{eqnarray}
where $T_{1(2)}= \omega_{1(2)}/{2\pi}$. 
{\color{black}  In Eq.~(\ref{eq:5}), it is assumed that $\hat{H}(w_1,w_2,t)$ comprised of two separate oscillatory terms such that the variable $t$ can be re-indexed to $t_{1}$ ($t_{2}$) on the term that oscillates with the frequency $\omega_{1}$ ($\omega_{2}$). Reversely, on Eq.~(\ref{eq:3}), we employ $t_{1}\!=\!t_{2} \!=\!t$, which indicates we intend to retrieve the time-dependent Hamiltonian, $\hat{H}(t_1,t_2)$, only on the line with $45^{\circ}$ in the continuous 2D time space. }
We then substitute the expansions given in Eqs.~(\ref{eq:3}) and (\ref{eq:4}) into the LvN equation [Eq.~(\ref{eq:1})] as 
\begin{eqnarray}
\label{eq:6}
\begin{aligned}
&
\sum_{mn}
\big(
\frac{d \hat{\rho}^{mn}(t)}{d t}+
i (n \omega_1+m \omega_2) \hat{\rho}^{mn}(t)
\big) 
e^{i n \omega_1 t}e^{i m \omega_2 t}=  
\\
&
-{i} \sum_{kl,k^{\prime}l^{\prime}} 
\big[ 
\hat{H}^{kl},\hat{\rho}^{k^{\prime}l^{\prime}}(t)
\big] 
e^{i(l+l^{\prime}) \omega_1 t}
e^{i(k+k^{\prime}) \omega_2 t}
= 
\\
&
-{i}\sum_{mn,k^{\prime}l^{\prime}}
\big[
\hat{H}^{(m-k^{\prime})(n-l^{\prime})}, \hat{\rho}^{k^{\prime}l^{\prime}}(t)
\big] 
e^{in\omega_1t}e^{im\omega_2t}.
\end{aligned}
\end{eqnarray}
In step (II), we should first define four new algebraic operators, $\hat{L}''_m$, $\hat{L}'_n$, $\hat{N}''$, and $\hat{N}'$, in the two-mode Fourier space. Note that in the single-mode Floquet theory, we define two algebraic operators which are called the \textit{Floquet Number}, $\hat{N}$, and the \textit{Floquet Ladder}, $\hat{L}_{n}$, operators \cite{mosallanejad2023floquet}. These operators have associations with a single-index Fourier {\color{black} basis}, $\{ |n\rangle \}$. 
In case of the two-mode Floquet theory, we need to build a two-index Fourier {\color{black} basis} by the tensor product of single-index Fourier {\color{black} basis} as $\{ |m,n\rangle \} \!=\! \{ |m \rangle \} \otimes \{ |n \rangle \}$ in which $\{ |n \rangle \}$ ($\{ |m \rangle \}$) corresponds to the first (second) frequency $ \omega_1$ ($ \omega_2$). 
{\color{black} The above four operators, in turn, are defined via the one-mode \text{Ladder} and \text{Number} operators as}
\begin{eqnarray}
\label{eq:7}
\begin{aligned}
&\hat{L}'_n = \hat{I}^{\omega_2} \otimes \hat{L}^{\omega_1}_n, \quad  
\hat{L}''_m = \hat{L}^{\omega_2}_m \otimes \hat{I}^{\omega_1}, \\
& \hat{N}' = \hat{I}^{\omega_2} \otimes \hat{N}^{\omega_1}, \quad 
\hat{N}'' = \hat{N}^{\omega_2} \otimes \hat{I}^{\omega_1} .
\end{aligned}
\end{eqnarray}
Here, $\hat{I}$ {\color{black} is} the identity operator. {\color{black}  The superscript indices ${\omega_{1}}$ and ${\omega_{2}}$} are used to distinguish the two Fourier spaces. These four operators obey the following properties 
\begin{eqnarray}
\label{eq:8}
\begin{aligned}
\hat{L}'_k|m, n\rangle = |m,  n+k\rangle,& \quad \hat{L}''_k|m, n\rangle = |m+k, n\rangle,\\
\hat{N}'|m, n\rangle = n|m, n\rangle,& \quad  
\hat{N}''|m, n\rangle = m|m, n\rangle,\\ 
 [\hat{N}', \hat{L}'_k] = k \hat{L}'_k,& \quad
[\hat{N}'', \hat{L}''_k] = k \hat{L}''_k,\\
[\hat{L}'_n, \hat{L}'_k] = 0,& \quad 
[\hat{L}''_m, \hat{L}''_k] = 0.
\end{aligned}
\end{eqnarray}
{\color{black}  $n$ and $m$ are harmonic indices that spans from negative to positive integer values.} Also, we should highlight that the single-primed operators act on the first index $n$, whereas the double-primed operators act on the second index $m$, of the two-mode Fourier {\color{black} basis}, in the same way as the original \text{Ladder} and \text{Number} operators act on the one-mode Fourier {\color{black} basis}. 
{\color{black}  One can show that operators in different subsets commute with each other as}
\begin{eqnarray}
\label{eq:9}
[\hat{N}^{\prime}, \hat{N}^{\prime\prime}] =0, \quad
[\hat{L}^{\prime}_{n}, \hat{L}^{\prime\prime}_{k}]=0, \quad
[\hat{L}^{\prime}_{n}, \hat{N}^{\prime\prime}]=0. \quad
\end{eqnarray} 
Next, we introduce the following Fourier representations
\begin{eqnarray}
\label{eq:10}
\hat{H}^{f}(t)&=&\sum_{mn} \hat{L}^{\prime\prime}_{m} \hat{L}^{\prime}_{n} \otimes \hat{H}^{mn}  
\, e^{i n \omega_1 t}e^{i m \omega_2 t}, \\
\label{eq:11}
\hat{\rho}^{f}(t)&=&\sum_{mn} \hat{L}^{\prime\prime}_{m} \hat{L}^{\prime}_{n}  
\otimes \hat{\rho}^{mn} (t)
\, e^{i n \omega_1 t}e^{i m \omega_2 t}.
\end{eqnarray} 
{\color{black} Here, we multiplied the Fourier coefficients in Eqs. (\ref{eq:3}) and (\ref{eq:4}) by the ladder operators $\hat{L}^{\prime\prime}_{m} \hat{L}^{\prime}_{n}=\hat{L}^{\omega_2}_{m} \otimes\hat{L}^{\omega_1}_{n}$.}
The Ladder operators transform the vector-like Fourier expansions into fairly complex matrix-like representations.
{\color{black} In the same manner as the single-mode Floquet theory,} we substitute Eqs. (\ref{eq:10}) and (\ref{eq:11}) in the LvN equation [Eq.~(\ref{eq:1})]. The RHS is given by
\begin{eqnarray}
\label{eq:12}
\begin{aligned}
\sum_{mn}
&
\hat{L}^{\prime\prime}_{m} \hat{L}^{\prime}_{n}
\otimes
\big(
\frac{d \hat{\rho}^{mn}(t)}{d t}+
i (n\omega_1+m\omega_2) \hat{\rho}^{mn}(t)
\big) \times ~~~~~~~~\\
&e^{i n \omega t}e^{i m \nu t}.
\end{aligned}
\end{eqnarray}
Temporarily ignoring the $-i$, and with the abbreviation $\hat{\rho}^{mn}\equiv \hat{\rho}^{mn}(t)$, the LHS is given by
\begin{eqnarray}
\label{eq:13}
\begin{aligned}
& \sum_{kl,k^{\prime}l^{\prime}} 
\big[ 
\hat{L}^{\prime\prime}_{k} \hat{L}^{\prime}_{l}
\otimes
\hat{H}^{kl} ,
\hat{L}^{\prime\prime}_{k^{\prime}} \hat{L}^{\prime}_{l^{\prime}}
\otimes
\hat{\rho}^{k^{\prime}l^{\prime}} 
\big] 
e^{i(l+l^{\prime})\omega_1 t}
e^{i(k+k^{\prime})\omega_2 t}=
\\& 
\sum_{mn,k^{\prime}l^{\prime}}
\big[
\hat{L}^{\prime\prime}_{m-k^{\prime}} 
\hat{L}^{\prime}_{n-l^{\prime}}
\otimes
\hat{H}^{(m-k^{\prime})(n-l^{\prime})}, 
\hat{L}^{\prime\prime}_{k^{\prime}} \hat{L}^{\prime}_{l^{\prime}}
\otimes 
\hat{\rho}^{k^{\prime}l^{\prime}}
\big]
\\&
\times e^{i n \omega_1 t}e^{i m \omega_2 t}=
\\&
\sum_{mn,k^{\prime}l^{\prime}}
\hat{L}^{\prime\prime}_{m} 
\hat{L}^{\prime}_{n}
\otimes
\big[
\hat{H}^{(m-k^{\prime})(n-l^{\prime})}
, 
\hat{\rho}^{k^{\prime}l^{\prime}}
\big]
e^{i n \omega_1 t}e^{i m \omega_2 t},
\end{aligned}
\end{eqnarray}
{\color{black}  where in the last line, we benefit from
$[\hat{L}^{\prime\prime (\prime)}_{m-k^{\prime}}, \hat{L}^{\prime\prime (\prime)}_{k^{\prime}} ] =0$.
Here, the main point is that Eqs. (\ref{eq:8}) and (\ref{eq:9}) allow us to factor out $\hat{L}^{\prime\prime}_{m}\hat{L}^{\prime}_{n}$.}
Combining Eq.~(\ref{eq:12}) and Eq.~(\ref{eq:13}), and recovering the $-i$, we arrive in  
\begin{eqnarray}
\label{eq:14}
\begin{aligned}
&\sum_{mn}
\hat{L}^{\prime\prime}_{m} 
\hat{L}^{\prime}_{n}
\!
\otimes
\!
\big(
\frac{d \hat{\rho}^{mn}}{d t}+
i (n\omega_1\!+\!m\omega_2) \hat{\rho}^{mn}
\big) 
e^{i n \omega_1 t}e^{i m \omega_2 t} =
\\
&-i\!\!\!\sum_{mn,k^{\prime}l^{\prime}}
\!\!\hat{L}^{\prime\prime}_{m} 
\hat{L}^{\prime}_{n}
\otimes
\big[
\hat{H}^{(m-k^{\prime})(n-l^{\prime})}
, 
\hat{\rho}^{k^{\prime}l^{\prime}}
\big]
e^{i n \omega_1 t}e^{i m \omega_2 t}.~~~~
\end{aligned}
\end{eqnarray}
Other than the operator $\hat{L}^{\prime\prime}_{m}\hat{L}^{\prime}_{n}$, the two sides of Eq.~(\ref{eq:14}) and Eq.~(\ref{eq:6}) are identical. Hence, we have proven that with proper definition for two-mode Floquet Number and Ladder operators, the LvN equation in Fourier representations keeps the original form as
\begin{eqnarray}
\label{eq:15}
\frac{d{\hat{\rho}}^f(t)}{dt} 
= -{i} [{\hat{H}}^f (t), {\hat{\rho}}^f(t)].
\end{eqnarray}
This is the end of step (II).

{\color{black}  In step (III), we desire to transform the LvN in Fourier representation [Eq.~(\ref{eq:15})] into the Floquet form by acting $e^{-i (\hat{N}^{\prime\prime} \omega_2+\hat{N}^{\prime} \omega_1) t}$ and $e^{i (\hat{N}^{\prime\prime} \omega_2+\hat{N}^{\prime} \omega_1) t}$ from the left and right sides, respectively. For ${\hat{\rho}}^f(t)$, it reads}
\begin{eqnarray} 
\label{eq:16}
\hat{\rho}^{F}(t)
&=&
e^{-i (\hat{N}^{\prime\prime} \omega_2+\hat{N}^{\prime} \omega_1) t} 
\hat{\rho}^{f} (t) 
e^{i (\hat{N}^{\prime\prime} \omega_2+\hat{N}^{\prime} \omega_1) t}
\\ \nonumber 
&=&
\sum_{mn}^{}
\hat{L}^{\prime\prime}_{m} 
\hat{L}^{\prime}_{n}
\otimes
\hat{\rho}^{mn}(t).~~~~~~~~~~~~~~
\end{eqnarray}
Notice that, to get the last line in the above expression, we have benefited from $[\hat{N}^{\prime},\hat{N}^{\prime\prime}]\!=\!0$ which allows us to shift the order by {\color{black} which} operators $\hat{N}^{\prime}$ and $\hat{N}^{\prime\prime}$ act on $\hat{\rho}^{f}$.
{\color{black}  Notably, similar to the single-mode Floquet theory, the commutation relations
$[\hat{N}^{\prime\prime}, \hat{L}^{\prime\prime}_{m}]\!=\!m \hat{L}^{\prime\prime}_{m}$ and $[\hat{N}^{\prime}, \hat{L}^{\prime}_{n}]\!=\!n \hat{L}^{\prime}_{n}$, along with the Baker-Campbell-Hausdorff (BCH) expansion, yield:
$e^{-i \hat{N}^{\prime\prime} \omega_2 t} \hat{L}^{\prime\prime}_{m} e^{i \hat{N}^{\prime\prime} \omega_2 t}\!=\!\hat{L}^{\prime\prime}_{m}e^{-i m \omega_2 t}$ and $e^{-i\hat{N}^{\prime} \omega_1 t} \hat{L}^{\prime}_{n} e^{i \hat{N}^{\prime} \omega_1 t}\!=\!\hat{L}^{\prime}_{n}e^{-i n \omega_1 t}$. These simplifications allow the exponential terms to cancel each other out.}
To proceed, we can inversely define $\hat{\rho}^{f}(t)$ in terms of $\hat{\rho}^{F}(t)$, and then obtain the time derivative of $\hat{\rho}^{f}(t)$ as
\begin{eqnarray} 
\label{eq:17}
\frac{d{\hat{\rho}^f}(t)}{dt} 
=
e^{i (\hat{N}^{\prime\prime} \omega_2+\hat{N}^{\prime} \omega_1) t}
\,
\frac{d{\hat{\rho}^F}(t)}{dt}
\,
e^{-i (\hat{N}^{\prime\prime} \omega_2+\hat{N}^{\prime} \omega_1) t}\, + \quad \quad \\ \nonumber
 e^{i (\hat{N}^{\prime\prime} \omega_2+\hat{N}^{\prime} \omega_1) t}
 \,
{i}
[\hat{N}^{\prime\prime} \omega_2+\hat{N}^{\prime} \omega_1, {\hat{\rho}}^F(t)]
\,
e^{-i (\hat{N}^{\prime\prime} \omega_2+\hat{N}^{\prime} \omega_1) t}.
\end{eqnarray}
Substituting the above relation into Eq.~(\ref{eq:15}) and applying the last transformation given in Eq.~(\ref{eq:16}) reads as
\begin{eqnarray}
\label{eq:18}
\frac{d{\hat{\rho}}^F}{dt} 
= -{i} [\sum_{mn} 
\hat{L}^{\prime\prime}_{m} 
\hat{L}^{\prime}_{n}
\otimes \hat{H}^{mn}
+\hat{N}^{\prime} \omega_1 +\hat{N}^{\prime\prime} \omega_2, {\hat{\rho}}^F].\quad ~~
\end{eqnarray}
Finally, we define the two-mode Floquet Hamiltonian as
\begin{eqnarray} 
\label{eq:19}
\hat{H}^{F}
= \sum_{mn}
\hat{L}^{\prime\prime}_{m} 
\hat{L}^{\prime}_{n}
\otimes
\hat{H}^{mn}
+\hat{N}^{\prime} \omega_1 +\hat{N}^{\prime\prime} \omega_2.
\end{eqnarray}
With the above definition for $\hat{H}^{F}$, Eq.~(\ref{eq:18}) is indeed the Eq.~(\ref{eq:2}), the two-mode Floquet LvN equation, which has the same structure as the traditional LvN. {\color{black}  Thus, the Eq.~(\ref{eq:2}) with $\hat{H}^{F}$ given in Eq.~(\ref{eq:19}) is exact, irrespective of the ratio between $\omega_1$ and $\omega_2$.}
\subsection{ \label{sec2_subsec:1} Time evolution and projection to the Hilbert space}
The time evolution associated with the Eq.~(\ref{eq:18}) is:
\begin{eqnarray}
\label{eq:20}
\hat{U}^F(t, t_0)=
e^{-i \hat{H}^F\left(t-t_0\right)}.
\end{eqnarray}
{\color{black}  Consequently, the Floquet density operator evolves as}
\begin{eqnarray}
\label{eq:21}
\begin{aligned}
\hat{\rho}^{F}(t)=\hat{U}^F(t, t_0)
\hat{\rho}^{F}(t_0) \hat{U}^{F^{-1}}(t, t_0).
\end{aligned}
\end{eqnarray}
{\color{black}  According to Eq.~(\ref{eq:3}), one requires to extract the coefficients $\hat{\rho}^{mn}(t)$ from $\hat{\rho}^F(t)$. To do this, one can use the properties $\langle k,q|\hat{L}^{\prime}_n\!=\!\langle k,q\!-\!n|$, and $\langle k,q|\hat{L}^{\prime\prime}_m\!=\!\langle k\!-\!m, q|$ as:}
\begin{eqnarray}
\label{eq:23}
\begin{aligned}
&
\!\!\!\sum_{mn}\langle k,q|
\hat{L}^{\prime\prime}_{m} 
\hat{L}^{\prime}_{n}
\otimes
\hat{\rho}^{mn}(t)
|0,0\rangle\!=\!\sum_{mn} 
\langle k\!-\!m,q\!-\!n|0,0\rangle  \\
&
\otimes \hat{\rho}^{mn}(t)=
\!\sum_{mn} \delta_{km} \delta_{qn} \otimes \hat{\rho}^{mn}(t)=\hat{\rho}^{kq}(t).
\end{aligned}
\end{eqnarray}
{\color{black}  Hence, $\hat{\rho}^F(t)$ projects to the Hilbert space as}
\begin{eqnarray}
\label{eq:24}
\begin{aligned}
\hat{\rho}(t)=
\sum_{mn}\langle m ,n| \hat{\rho}^F (t)
|0,0\rangle 
e^{i n \omega_1 t} 
e^{i m \omega_2 t}.
\end{aligned}
\end{eqnarray}
Such projection can be applied to the $\hat{U}^F(t,t_0)$ to arrive at a time evolution operator in the Hilbert space as
\begin{eqnarray}
\label{eq:25}
\begin{aligned}
&\hat{U}(t, t_0)=
\sum_{mn}\langle m, n|e^{-i \hat{H}^F\left(t-t_0\right)}|0,0\rangle 
e^{i n \omega_1 t}
e^{i m \omega_2 t}.
\end{aligned}
\end{eqnarray}
{\color{black}  Eq.~(\ref{eq:25}) indeed satisfies the initial condition $\hat{U}(t_0,t_0)\!=\!\hat{I}$. We will adopt this time-evolution expression for open quantum systems, as will be explained shortly, even though the two-mode driven Hamiltonian is not strictly periodic.}

\subsection{\label{sec2_subsec:2} Electronic Model Hamiltonian} 
\begin{figure}[h]
	\begin{center}
	\includegraphics[width=5.0cm]{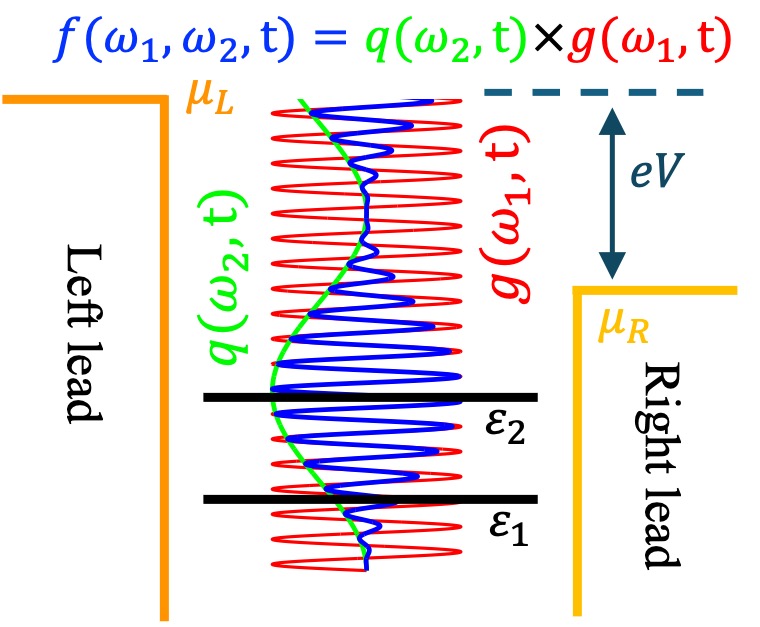}
	\end{center}	\caption{\label{fig:1scheme} Schematic setup for quantum transport through a multi-level mesoscopic system driven simultaneously by two periodic oscillations.}
\end{figure}

We consider a multi-level system (dot) driven by two periodic frequencies, either through on-site or off-diagonal coupling, as depicted in Fig.~(\ref{fig:1scheme}). {\color{black} The dot} can be driven by the simultaneous presence of two time-periodic external fields with different frequencies, $\omega_1$ and $\omega_2$. The multi-level system is also weakly coupled to the left (L) and right (R) electron baths. The electrons in the leads are assumed to have no interactions with each other, representing an ideal bath in thermal equilibrium. The spinless Hamiltonian of our model system is given by:
\begin{eqnarray}
\label{eq:26}
\hat{H}_{}(t) &=& \hat{H}_S(t)  + \hat{H}_B + \hat{H}_{SB}  
\\
\label{eq:27}
\hat{H}_S(t) &=& \sum_{ij} h_{ij} (\omega_1,\omega_2, t) \hat{d}_i^\dagger \hat{d}_j \\
\label{eq:28}
\hat{H}_B &=& \sum_{\alpha k} \epsilon_{\alpha k} \hat{c}_{\alpha k}^\dagger \hat{c}_{\alpha k} 
\\
\label{eq:29}
\hat{H}_{SB}  &=&  \sum_{\alpha k,i} 
V_{\alpha k, i}  
\hat{c}_{\alpha k}^\dagger \hat{d}_i +\mathrm{H.c.} = \sum_{\alpha i}  
\hat{C}_{\alpha i}^\dagger \hat{d}_i +\mathrm{H.c.}~~~~~
\end{eqnarray}
Without time-dependent driving, the above Hamiltonian system is known as the multi-level Anderson model~\cite{ryndyk2016theory}.
Here, we consider the dot that is strongly driven by two independent functions, oscillating at two different frequencies, such that perturbative methods fail for both components. The total system Hamiltonian is not perfectly periodic, except when the ratio between the frequencies equals to the ratio of two integers, $\omega_1/\omega_2\!=\!n_1/n_2$. The bath Hamiltonian, $\hat{H}_{B}$, and the system-bath interaction, $\hat{H}_{SB}$, are assumed to remain time-independent.
Here, $\hat{d}_i$ ($\hat{d}_i^{\,\dagger}$) is the dot's many-body electronic annihilation (creation) operator, and $h_{ij}(\omega_1,\omega_2,t)$ represents the one-body Hamiltonian driven by the two frequencies.
Similarly, $\hat{c}_{\alpha k}$ ($\hat{c}_{\alpha  k}^\dagger$) is the annihilation (creation) operator for the $k$th electronic level in the bath $\alpha\in L,R$.
The $V_{\alpha k,i}$ determines the coupling strength between the $k$th orbital on the bath $\alpha$, and $i$th orbital level in the system.
Note that in the second part of Eq.~(\ref{eq:29}), we have re-expressed the interaction Hamiltonian based on definition $\hat{C}_{\alpha i}^\dagger=\sum_{k} V_{\alpha k, i}\hat{c}_{\alpha k}^\dagger$ to simplify the notation of the system-bath interaction. The bath $\alpha$ is also associated with the electrochemical potential $\mu_{\alpha}$, which determines its statistical properties. Here, we have presented the spinless Hamiltonian for the sake of compactness and clarity. However, as explained in Section III, extending the spinless study to its spinful counterpart is a straightforward process.
\subsection{ \label{sec2_subsec:3} Wangsness-Bloch-Redfield (WBR) QME}
{\color{black} 
Our theoretical formulation of an open quantum system starts with the Wangsness-Bloch-Redfield (WBR) QME as \cite{timm2008tunneling, mosallanejad2025floquet}:}
\begin{eqnarray}
\begin{aligned}
\label{eq:35}
\frac{d\hat{\rho}_S(t)}{dt}= 
-i [\hat{H}_{S}(t), &~\hat{\rho}_S(t)]  
-\!\int_{0}^{\infty} \!\!
d\tau
\operatorname{Tr}_B [\hat{H}_{SB},\\
&[\hat{\tilde{H}}_{SB}(\tau,t), \hat{\rho}_S(t) \otimes\hat{\rho}_B ] ].
\end{aligned}
\end{eqnarray}
{\color{black}  The simplified key operator $\hat{\tilde{H}}_{SB}(\tau,t)$ is given by}
\begin{eqnarray}
\begin{aligned}
\label{eq:36}
\hat{\tilde{H}}_{SB}(\tau,t)\!=\! 
U_B(\tau)U_S(t,t\!-\!\tau)  
\hat{H}_{SB} 
U_S^{\dagger}(t,t\!-\!\tau) U_B^{\dagger}(\tau),~~~~
\end{aligned}
\end{eqnarray}
{\color{black}  with $U_S(t,t\!-\!\tau)=\mathcal{T} e^{-i \int_{t-\tau}^t\hat{H}_S(s)ds}$ being the two time evolution operators relevant to the system only, and $U_B(\tau)\!=\!e^{-i\hat{H}_B\tau}$ being the same for the bath. Here, $\mathcal{T}$ is the time-ordering operator.}
In addition, if we take $H_S$ time-independent{\color{black} , above} WBR QME reduces to a well-accepted form in which the evolution of $\hat{\tilde{H}}_{SB}$ only determines with the time difference $\tau$~\cite{elste2005theory}. 
{\color{black} Under accepted approximations, Eq.~(\ref{eq:35}) is highly general and can be applied to any time-dependent Hamiltonian governing the restricted system (dot).} Eq.~(\ref{eq:35}) requires significant simplifications before it can be used practically. Following cases in which system's Hamiltonian is time-independent, we are expecting to arrive at explicit forms for dissipative terms, so-called \textit{dissipators}. 
{\color{black}  Eq.~(\ref{eq:36}) reflects a important observation: the possible dissipators would become time dependent because $\hat{\tilde{H}}_{SB}$ itself is time dependent. Deriving practical expressions for the dissipators involves three stages,} as will be detailed in the following three subsections. In the first stage, we trace out the bath degrees of freedom. In the second stage, we perform the time integration over $\tau$. Finally, in the third stage, we apply the wide-band approximation to justify the final form of the dissipators.
\subsection{ \label{sec2_subsec:4} Time-dependent dissipators for QME - I: tracing out bath}
{\color{black} Here, we will disentangle the double commutation in Eq.~(\ref{eq:35}) with substituting explicit forms of $\hat{H}_{SB}$ and $\hat{\tilde{H}}_{SB}$. To do this transparently, we redefine $\hat{\tilde{H}}_{SB}$ as} 
\begin{eqnarray}
\begin{aligned}
\label{eq:38}
\hat{\tilde{H}}_{SB}(\tau,t)=\sum_{li} \tilde{C}_{\alpha i}^{\dagger}(\tau) \tilde{d}_{i}(t, \tau) 
+ \tilde{d}_{i}^{\dagger}(t, \tau) \tilde{C}_{\alpha i}(\tau),~~~~~~
\end{aligned}
\end{eqnarray}
{\color{black}  with $\tilde{C}_{\alpha i}(\tau)\!=\! U_B(\tau) \hat{C}_{\alpha i}U_B^{\dagger}(\tau)\!=\!\sum_{\alpha k} V_{\alpha k,i}^{\ast}\hat{c}_{\alpha k} e^{i\,\epsilon_{\alpha k}\tau}$, and $\tilde{d}_{i}^{\,(\dagger)}(t, \tau)\!=\!U_S(t,t\!-\!\tau)\hat{d}_{i}^{\,(\dagger)}
U_S^{\dagger}(t,t\!-\!\tau)$.}
{\color{black} The structure of $\tilde{C}_{\alpha i}(\tau)$ can be understood via the combined use of the fermionic Pauli principle and the BCH formula~\cite{breuer2002theory}.}
Note that $\tilde{C}^{\dagger}_{\alpha i}(\tau)\!=\!\tilde{C}_{\alpha i}(\tau)^{\, \dagger}$ and $\tilde{d}^{\dagger}_{i}(t, \tau)\!=\!\tilde{d}_{i}(t, \tau) ^{\, \dagger}$.  
Disentangling the double commutation yields four generic integrands. The first and second generic integrands read as: $\hat{H}_{SB} \hat{\tilde{H}}_{SB}(\tau,t) \hat{\rho}_S(t) \otimes \hat{\rho}_B$, and $\hat{H}_{SB} \hat{\rho}_S(t)\! \otimes \! \hat{\rho}_B \hat{\tilde{H}}_{SB}(\tau,t)$. Each of these four generic integrands comprises four specific integrands. 
{\color{black} 
When faced with the first generic integrand, and with further rearrangements, we only keep following two specific terms}
\begin{eqnarray}
\begin{aligned}
\label{eq:41}
\sum_{\smash{\alpha i, \beta j}} 
\,
&\hat{C}_{\alpha i}^{\dagger}
\tilde{C}_{\beta j}(\tau) 
\hat{\rho}_B
\otimes 
\hat{d}_{i}
\tilde{d}_{j}^{\dagger}(t, \tau) 
\hat{\rho}_S(t)+\\[-6pt]
&\hat{C}_{\alpha i}
\tilde{C}_{\beta j}^{\dagger}(\tau)
\hat{\rho}_B 
\otimes 
\hat{d}_{i}^{\dagger} 
\tilde{d}_{j}(t, \tau) 
\hat{\rho}_S(t).
\end{aligned}
\end{eqnarray}
{\color{black} The terms where $\hat{C}$ [$\hat{C}^{\dagger}$] multiplies $\tilde{C}$ [$\tilde{C}^{\dagger}$] are discarded, as their trace vanishes.}
We refer to the above two terms as the final first and second terms. Similarly, the final third and fourth non-vanishing integrands are 
\begin{eqnarray}
\begin{aligned}
\label{eq:42}
\sum_{\smash{\alpha i, \beta j}} 
\,
&\hat{C}_{\alpha i}^{\dagger}
\hat{\rho}_B
\tilde{C}_{\beta j}(\tau) 
\otimes 
\hat{d}_{i}
\hat{\rho}_S(t) 
\tilde{d}_{j}^{\dagger}(t, \tau)+\\[-6pt]
&\hat{C}_{ \alpha i}
\hat{\rho}_B 
\tilde{C}_{\beta j}^{\dagger}(\tau)
\otimes 
\hat{d}_{i}^{\dagger} 
\hat{\rho}_S(t)
\tilde{d}_{j}(t, \tau).
\end{aligned}
\end{eqnarray}
{\color{black}  Using the explicit form of the operators $\hat{C}_{\alpha i}^{\dagger}$ and $\tilde{C}_{\alpha i} (\tau)$ 
[$\hat{C}_{\alpha i}$ and $\tilde{C}_{\alpha i}^{\dagger} (\tau)$]
in terms of $\hat{c}_{\alpha k}$ and $\hat{c}_{\alpha k}^{\dagger}$, and the bath's density
operator in the first [second] term, we are now able to trace out the bath degrees of freedom as: $\operatorname{Tr}_B (\hat{c}_{\alpha k}^{\dagger} \hat{c}_{\beta k^\prime} \hat{\rho}_B(\mu) )\!=\!f(\epsilon_{\alpha k}, \mu_{\alpha})\delta_{k,k^\prime}\delta_{\alpha,\beta}$
[$\operatorname{Tr}_B( \hat{c}_{\alpha k} \hat{c}_{\beta k^\prime}^{\dagger} \hat{\rho}_B(\mu) )\!=\!1\!-\!f(\epsilon_{\alpha k},
\mu_{\alpha}) \delta_{k,k^\prime}\delta_{\alpha,\beta}$]~\cite{pourfath2014non}.
The process of tracing out the bath degrees of freedom in the third and fourth terms is carried out in the same way. However, one has to be cautious with the third and fourth terms, as the cyclic invariance of the trace must be utilized.}
Thereupon, tracing out the bath degrees of freedom gives us the following expression for the first four non-vanishing terms (out of eight)
\begin{eqnarray}
\begin{aligned}
\label{eq:43}
& \int_{0}^{\infty} 
\!\!
d\tau
\!\!
\sum_{\smash{\alpha k, i j}}
\!
V_{\alpha k, i}
V_{\alpha k,j}^{\ast}
f(\epsilon_{\alpha k}, \mu_{\alpha})
e^{i \, \epsilon_{\alpha k}  \tau}
\,
\hat{d}_{i}
\tilde{d}_{j}^{\dagger}(t, \tau) 
\hat{\rho}_S(t)+ \\[-4pt]
&V_{\alpha k, i}^{\ast}
V_{\alpha k,j}
(1\!-\!f(\epsilon_{\alpha k}, \mu_{\alpha}))
e^{-i \, \epsilon_{\alpha k}  \tau}
\,
\hat{d}_{i}^{\dagger} 
\tilde{d}_{j}(t, \tau) 
\hat{\rho}_S(t)
- ~~\\  
&V_{\alpha k, i}
V_{\alpha k,j}^{\ast}
(1\!-\!f(\epsilon_{\alpha k}, \mu_{\alpha}))
e^{i \, \epsilon_{\alpha k}  \tau}
\,
\hat{d}_{i}
\hat{\rho}_S(t)
\tilde{d}_{j}^{\dagger}(t, \tau) 
- ~~ \\ 
&V_{\alpha k, i}^{\ast}
V_{\alpha k,j}
f(\epsilon_{\alpha k}, \mu_{\alpha})
e^{-i \, \epsilon_{\alpha k}  \tau~~}
\,
\hat{d}_{i}^{\dagger} 
\hat{\rho}_S(t)
\tilde{d}_{j}(t,\tau). 
\end{aligned}
\end{eqnarray}
{\color{black} Notice that the double summation of the indices $k$ and $k^\prime$ reduces to only one summation, namely $k$. The same reduction in the summation index also applies to indices $\alpha$ and $\beta$.}
For the sake of clarity and brevity, we retain only the first four integrands of the full set of eight integrands, as it provides a sufficient description for the process of simplification of dissipators.
{\color{black}  Here, we highlight that performing the integration analytically is infeasible until the explicit $\tau$-dependence of $\tilde{d}_{i/j}$, and $\tilde{d}_{i/j}^{\,\dagger}$ is known.}
\subsection{ \label{sec2_subsec:5} 
{\color{black} 
Time-dependent dissipators for QME - II: integration over $\tau$} }
{\color{black}  A clear distinction between the time variables $t$ and $\tau$ is essential.} In practice, we first need to identify the explicit form of $U_S(t,t-\tau)$, and then define $\tilde{d}_{j}(t, \tau)$ and $\tilde{d}^{\dagger}_{j}(t, \tau)$ as functions of $t$ and $\tau$. In general, when the Hamiltonian of the system is time-dependent, deriving an analytical form for the time-evolution operator is not a trivial task, primarily because $H_S(t)$ does not necessarily commute with itself at different times~\cite{gerry2023introductory}. {\color{black}  Floquet theory provides an explicit form for the time-evolution operator when the Hamiltonian is exactly periodic, known as Shirley's time-evolution formula~\cite{shirley1965solution}.} In our previous work, we showed that this form is advantageous for deriving a time-dependent dissipator. As demonstrated in \autoref{sec2_subsec:0} and \autoref{sec2_subsec:1}, the theory can be extended to {\color{black}  two-mode Floquet theory}. {\color{black}  Using Eq.~(\ref{eq:25}), we can re-express $\tilde{d}_{j}(t,\tau)$ and $\tilde{d}_{j}^{\,(\dagger)}(t, \tau)$
\begin{eqnarray}
\label{eq:45}
\begin{aligned}
&\tilde{d}_{i}^{\,(\dagger)}(t,\tau)
\!=\!
U_S(t,t\!-\!\tau)\hat{d}_{i}^{\,(\dagger)}
U_S^{\dagger}(t,t\!-\!\tau)\!=\!
\sum_{mn,kq} 
\\& 
\langle mn|\hat{Y} 
e^{-i \hat{\Lambda}_S^F \tau}
\hat{\mathcal{D}}_i^{o\,(\dagger)}
e^{i \hat{\Lambda}_S^F \tau}
\hat{Y}^{\dagger} |kq\rangle 
e^{i (n-q) \omega_1 t} e^{i (m-k) \omega_2 t},
\end{aligned}
\end{eqnarray}
in which we have defined $\hat{ \mathcal{D}}_i^{o\,(\dagger)}\!\!=\!\! \hat{Y}^{\dagger} |00\rangle \hat{d}^{(\dagger)}_{i} \langle 00|\hat{Y}$, and the two-mode Floquet Hamiltonian diagonalized as $\hat{\Lambda}_S^F=\hat{Y}^{\dagger}\hat{H}_S^F\hat{Y}$.}
Now, the distinction between $\tau$ and $t$ becomes evident in Eq.~(\ref{eq:45}). 
{\color{black} We must then choose a Hilbert space basis}, $\{|a\rangle\}$, so that operators can be expressed as matrices. For instance, the elements of the density operator can be written as $\rho_{ab}(t) \!=\! \langle a|\hat{\rho}(t)|b \rangle$. By reducing operators to matrices, particularly the rotating matrix $Y$, we can work within the diagonalized two-mode Floquet space, $\{|\gamma\rangle\}$, such that $\langle\gamma| \hat{\Lambda}_S^F\!=\!\langle\gamma|\mathcal{E}_{\gamma}$.
This allows us to express the matrix elements of the central operator within the bracket in Eq.~(\ref{eq:45}) as
\begin{eqnarray}
\begin{aligned}
\label{eq:46}
\left( 
e^{-i \hat{\Lambda}^F \tau} 
\hat{D}_j^{o (\dagger)} 
e^{i \hat{\Lambda}^F \tau} 
\right)_{\gamma \nu} 
= e^{-i \Omega_{\gamma \nu} \tau} \left( \hat{D}_j^{o (\dagger)} \right)_{\gamma \nu},
\end{aligned}
\end{eqnarray}
where $\Omega_{\gamma\nu}\!=\!\mathcal{E}_{\gamma}\!-\!\mathcal{E}_{\nu}$. 
Note that, we combined the two oppositely signed exponential terms into a single expression. 
{\color{black} Now, we can perform integration over $\tau$ as}
\begin{eqnarray}
\begin{aligned}
\label{eq:47}
\int_{0}^{\infty} 
\!\!\!\!
e^{\pm i(\epsilon_{\alpha k} \mp \Omega_{\gamma \nu}) \tau} \! d\tau 
\!=\!
\pi \delta(\epsilon_{\alpha k}
\!\mp \!
\Omega_{\gamma \nu}) 
\!
\pm
\!
i \mathcal{P}
(\!
\frac{1}{\epsilon_{\alpha k}\!\mp \!\Omega_{\gamma \nu}}
\!),~~~~
\end{aligned}
\end{eqnarray}
{\color{black} 
in terms of the delta function and the Cauchy's principal value.} Hereafter, we neglect the Cauchy’s principle term.
\subsection{ \label{sec2_subsec:6} {\color{black} Time-dependent} dissipators for QME - III: final step}
At this point, before substituting $\tilde{d}^{\dagger}$ and $\tilde{d}$ into Eq.~(\ref{eq:43}), we shall invoke the wide-band approximation. In general, when the electronic levels in the bath are very closely spaced, we can replace the summation, over $k$, by an integral as{\color{black} : $\sum_{\alpha k} \!\rightarrow \!\sum_{\alpha} \int D^{\alpha} (\epsilon_{\alpha k}) d\epsilon_{\alpha k}$.}
The wide-band approximation assumes that the DOS of an ideal bath, $D^{\alpha}$, does not change significantly over the range of energies relevant to the transport such that {\color{black}  $\Gamma_{i j}^{\alpha} \!=\! 2 \pi V_{\alpha k,i} V_{\alpha k,j}^* D^{\alpha} \!\approx\! \text{constant}$}.
$\Gamma_{ij}^{\alpha}$ is the coupling rate of the bath $\alpha$. We can perform the integration over $\epsilon_{\alpha k}$ in which the Fermi function essentially picks out the element $\Omega_{\gamma \nu}$. 
{\color{black}  Consequently}, the first four lead-specific dissipators (relevant to the bath $\alpha$) can be simplified as 
\begin{eqnarray}
\begin{aligned}
\label{eq:50}
\mathfrak{D}_{\alpha}^{I-IV}
\!=\!
\sum_{\smash{ i j}}  
& \frac{ \Gamma_{ij}^{\alpha} }{2}
{d}_{i} 
\mathbb{d}_{\alpha,j}^{\dagger\, + }(t)
{\rho}_S(t)  
+
\frac{ \Gamma_{ji}^{\alpha} }{2}
{d}_{i}^{\, \dagger} 
\thickbar{\mathbb{d}}_{\alpha,j}^{\,\,\, - }(t)
{\rho}_S(t) ~~~~~\\
-& \frac{ \Gamma_{ij}^{\alpha} }{2}
{d}_{i} 
{\rho}_S(t) 
\thickbar{\mathbb{d}}_{\alpha,j}^{\dagger\, + }(t)
-
\frac{ \Gamma_{ji}^{\alpha} }{2}
{d}_{i}^{\, \dagger} 
{\rho}_S(t) 
\mathbb{d}_{\alpha,j}^{\,\,\,- }(t),
\end{aligned}
\end{eqnarray}
where we have defined new time-dependent matrices 
\begin{eqnarray}
\label{eq:51}
\begin{aligned}
\mathbb{d}_{\alpha,j}^{(\dagger)\, \pm}(t)
\!=\!\!
\sum_{mn,kq}
& 
\!\!
\langle mn|
Y
f(\pm\Omega, \mu_{\alpha})
\circ
\mathcal{D}_j^{o \, (\dagger)}
Y^{\dagger} 
|kq\rangle \times \\
& 
e^{i (n-q) \omega_1 t}e^{i (m-k) \omega_2 t}, \\
\thickbar{\mathbb{d}}_{\alpha,j}^{(\dagger)\, \mp }(t)
\!=\!\!
\sum_{mn,kq}
\!\!
& 
\langle mn|
Y
(1\!-\!f(\mp\Omega, \mu_{\alpha}))
\circ
\mathcal{D}_j^{o \, (\dagger)}
Y^{\dagger} 
|kq\rangle \times \quad \\
\,& 
e^{i (n-q) \omega_1 t}e^{i (m-k) \omega_2 t}.
\end{aligned}
\end{eqnarray}
Note that in Eq.~(\ref{eq:51}), there are two types of occupation matrices: electron occupation, $f(\pm\Omega, \mu_{\alpha})$, and hole occupation, $1\!-\!f(\mp\Omega, \mu_{\alpha})$. These occupation matrices multiply to $\mathcal{D}_j^{o (\dagger)}$ by the Hadamard product, remarked by $\circ$. 
Accordingly, one can follow the same procedure for the second four non-vanishing dissipators, $\mathfrak{D}_{\alpha}^{V-VIII} ({\rho}_S(t),t )$.
However, the second set of four non-vanishing dissipators is the Hermitian conjugates of the first set. This symmetry can be utilized to reduce the computational cost of calculations. Finally, we can have the following compact dynamical equation for the reduced density matrix, $\rho_S(t)$, as
\begin{eqnarray}
\begin{aligned}
\label{eq:52}
\frac{d\rho_S(t)}{dt}= 
-i [H_{S}(t), \rho_S(t)]  
- \sum_{\alpha}
\mathfrak{D}_{\alpha}^{I-VIII} ({\rho}_S(t),t ). ~~~~
\end{aligned}
\end{eqnarray}
{\color{black}  One can straightforwardly extend the spinless quantum transport theory to include spin degrees of freedom.} First, annihilation (creation) operators must receive the spin degrees of freedom, $\sigma \in \{\uparrow,\downarrow\}$, as $\hat{d}_{i,\sigma}$ ($\hat{d}_{i,\sigma}^{\,\dagger}$). Consequently, we should define $\hat{ \mathcal{D}}_{i,\sigma}^{o}\!=\! \hat{Y}^{\dagger} |00\rangle \hat{d}_{i,\sigma} \langle 00|\hat{Y}$ and so on. This essentially indicates that all matrix operators in Eq.~(\ref{eq:50}) and Eq.~(\ref{eq:51}) should be decorated with the spin degrees of freedom such that one should consider both the spin-up and spin-down dissipators.      
\subsection{ \label{sec2_subsec:7} Observables }
Essentially, we need to define the system's electronic number operator in the Hilbert space as $\hat{n}_S=\sum_{i} \hat{d}_i^\dagger\hat{d}_i$. With that, we can evaluate the expectation of the system's particle number, $\langle \hat{n}_S \rangle(t)$ (known also as the total occupation or the charge number). Similarly, {\color{black}  the charge} number operator for each lead is defined as $\hat{n}_{\alpha}\!=\!\sum_{k} \hat{c}_{\alpha k}^\dagger\hat{c}_{\alpha k}$, where $\alpha \in \{L, R\}$. In general, the total charge number of the system at time t is given by
\begin{eqnarray}
\begin{aligned}
\label{eq:53}
\langle \hat{n}_S \rangle(t)= 
\operatorname{Tr} (\hat{n}_S\hat{\rho}_S(t)).
\end{aligned}
\end{eqnarray}
The second observable is the terminal current. The particle current operator that passes through the terminal $\alpha$, is defined as the rate of change of the total number of particles in that lead as: ${d \langle \hat{n}_{\alpha} \rangle(t)}/{dt}={d (\hat{n}_{\alpha}\hat{\rho}(t))}/{dt}$. Based on particle conservation, one can perceive that the rate of particle change in the reduced system is the sum of the currents passing through all terminals. 
{\color{black}  Hence, the charge current, $\hat{I}_{\alpha}$, can reversely be defined as} 
\begin{eqnarray}
{\color{black}  
\begin{aligned}
\label{eq:54}
&e \frac{d \langle \hat{n}_{S} \rangle(\bar{t})}{d\bar{t}}
=
\frac{e}{\hbar} 
\operatorname{Tr}  \big(
 \hat{n}_S \frac{d \hat{\rho}_S(t)}{dt}
 \big)= \\
&  
\!
\frac{e}{\hbar} 
\sum_{\alpha}
\operatorname{Tr} 
\big(
\hat{n}_S
\,
\hat{\mathfrak{D}}_{\alpha}^{I-VIII} (\hat{\rho}_S(t),t)
\big)\!=\! \sum_{\alpha} 
\hat{I}_{\alpha}(\bar{t}),~~~~
\end{aligned}
}
\end{eqnarray}
{\color{black}  where $e$ is the elementary charge. In above,} we make use of Eq.~(\ref{eq:52}) and $\operatorname{Tr} (\hat{n}_S [\hat{H}_{S}(t),\hat{\rho}_S(t)])\!=\!0$ due to the cyclic invariance of the trace and the fact that for fermions $[\hat{H}_{S},\hat{n}_S ]\!=\!0$. 
{\color{black}  Notice how the real current, $\hat{I}_{\alpha} (\bar{t})$ is related to the scaled particle current, $\operatorname{Tr} \big(\hat{n}_S\hat{\mathfrak{D}}_{\alpha}^{I-VIII} (t)\big)$, by the factor $e/\hbar$.} 
Here, with regard to the static aspect of transport, we cannot properly define the concept of a steady-state observable in the long-time limit for scenarios involving two periodic drivings, as the off-diagonal coupling is not necessarily perfectly periodic. This means that the system observable in the long-time limit may not settle to a steady constant value, instead it may oscillate unpredictably. Nonetheless, one can use the following definition for a quasi-steady state observable, 
\begin{eqnarray}
\label{eq:55}
{\color{black} 
{O}^{q}= \lim_{t\to\infty}  \frac{1}{\mathscr{T}} 
\int_{t}^{\,t+\mathscr{T}}  
\!dt \, {O}(t),
}
\end{eqnarray}
where $\mathscr{T}$ indicates the period by which the observable oscillates. {\color{black}  When including spin degrees of freedom, one must employ spin-resolved number operators} $\hat{n}^{\sigma}_S=\sum_{i}\hat{d}_{i,\sigma}^\dagger\hat{d}_{i,\sigma}$ in Eqs.~(\ref{eq:53}) and (\ref{eq:54}) to distinguish between spin-up and spin-down observables.
\section{Representative Applications}
\label{sec:3}
\subsection{\label{sec3_subsec:1} Transport through a spinless two-level dot}
\begin{figure}[h]
	\begin{center}	\includegraphics[width=7.8cm]{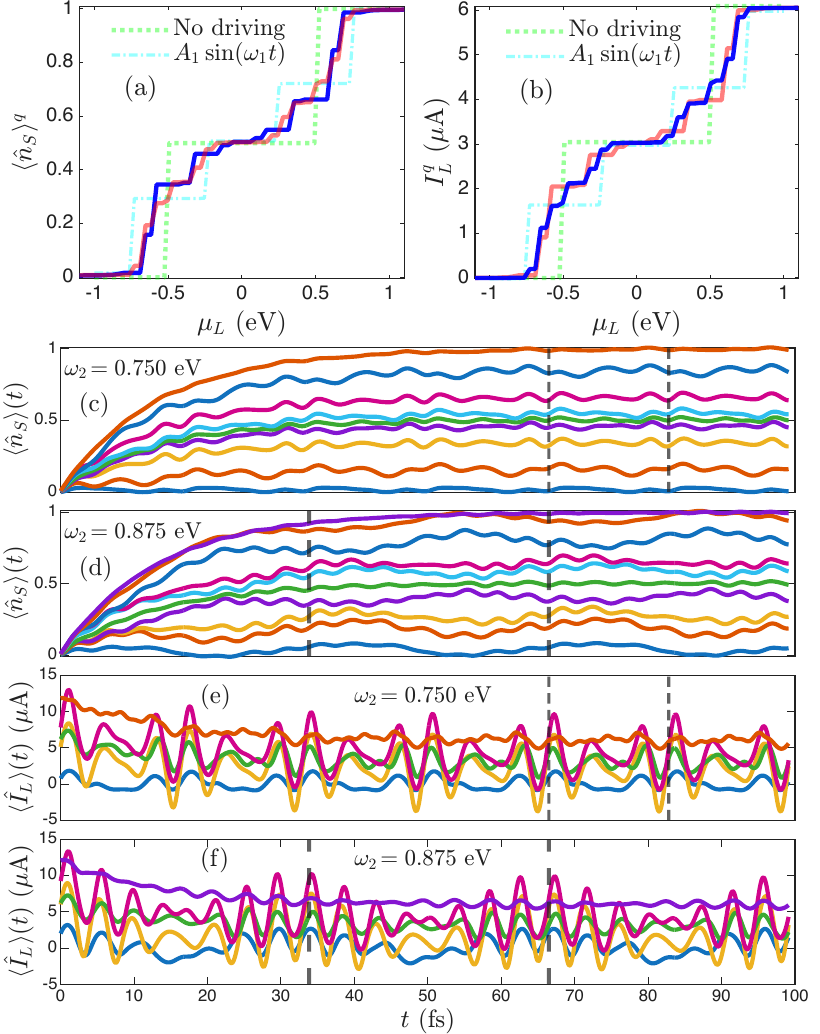}
	\end{center}
	\caption{\label{fig:2} (color online). {\color{black} Quasi-steady} and dynamics of an open two-level dot driven by two additive drivings. {\color{black} (a) Quasi-steady charge number. (b) Quasi-steady left-current}. (c)-(f) The corresponding time-dependent charge number and {\color{black} left-current}.}
\end{figure}

As preliminary applications of Eqs.~(\ref{eq:50})-(\ref{eq:55}), let us to consider electron transport through a spinless two-level quantum dot driven by two simultaneous off-diagonal couplings: (i) a primary coupling $g(\omega_1,t)$, and (ii) a secondary coupling $q(\omega_2,t)$. The general one-body Hamiltonian matrix of the system can be expressed as
\begin{eqnarray}
\label{eq:57}
h(\omega_1,\omega_2,t)\!=\!\!
\begin{bmatrix}
\epsilon_1 &\! g(\omega_1,t)\!\odot\!q(\omega_2, t) \\
g^*(\omega_1,t)\!\odot\!q^*(\omega_2,t) &\! \epsilon_2 
\end{bmatrix}\!,~~~~~
\end{eqnarray}
where $\odot\!\in\!\{+,\times\}$. 
Here, the goal is to inspect how incorporating a secondary driving term within different scenarios will alter the transport characteristics. Both the time-dependent and quasi-steady-state characteristics will be investigated. We recall that the current-\textit{voltage} ($\mu_L\!-\!\mu_R$) characteristic of a non-driven non-interacting multi-level system shows the typical step-like increase (conductance step) as the bias voltage increases, where each step corresponds to involving a new level into the transport, see doted lines in Figs.~\ref{fig:2}(a) and \ref{fig:2}(b). In our previous work, we observed that a cosine/sine off-diagonal coupling [e.g., $h_{12/21}(t)\!=\!A_1\sin(\omega_1 t)$] gives rise to the doubling of conductance steps when the coupling strength, $A_1$, is large compared to the {\color{black} energy gap} between the two levels, $\Delta \epsilon\!=\!\epsilon_2\!-\!\epsilon_1$. It is observed that the largest changes in conductance steps occur in the strong coupling regime when the driving frequency resonates with the {\color{black} energy gap}, $\omega_1\!=\!\Delta \epsilon$, and $A_1\!=\!\Delta \epsilon/2$; {\color{black}  see the dot-dashed lines} in Figs.~\ref{fig:2}(a) and \ref{fig:2}(b). Here, we keep the simulation temperature at 4.2 K (equivalent to the temperature energy $1/\beta\!\approx\!0.36$ meV) because in the low temperature regime, the thermal energy does not cause significant smearing of the Fermi distribution, leading to sharp conductance steps, and hence driven induced effects can be more pronounced. We set $\epsilon_{1}\!=\!-0.5$ eV, $\epsilon_{2}\!=\!+0.5$ eV, and the left and right electrodes are parameterized by the coupling rate of $\Gamma_{ji}^{L/R}\!=\!0.025$ eV. Hereafter, $\mu_R$ (drain) is fixed at a negative minimum voltage, $V_{min}$, far below $\epsilon_{1}$, while $\mu_L$ sweeps within the range [$V_{min}$,$V_{max}$].

\emph{ 1. Additive term scenario:}
In the first set of examples, we consider adding a secondary off-resonance deriving term to the main in-resonance driving, $\odot\!=\!+$. The primary driving term is $g(\omega_1,t)=A_1 \sin(\omega_1 t)$, and the secondary driving term is $q(\omega_2,t)=A_2 \sin(\omega_2 t)$. This two-level system with two off-diagonal drivings represents a molecular junction under dual linearly polarized laser illumination, where both lasers couple to electronic levels through dipole interactions. Here, $A_{1/2}$ represents the transition dipole moment $\mu$ multiplied by the amplitude of the laser's electric field $E_0$~\cite{peskin2012coherently}. The main driving frequency is fixed at $\omega_1\!=\!1.0$ eV(in-resonance, mid-infrared laser) and the driving amplitudes are set equal to $A_{1/2}\!=\!0.25$ eV. In Figs.~\ref{fig:2}(a) and \ref{fig:2}(b), we present the {\color{black} quasi-steady} particle number and the {\color{black} quasi-steady} current at the left terminal (left-current) for two values of the secondary frequency: $\omega_2\!=\!0.750$ eV and $\omega_2\!=\!0.875$ eV (off-resonance).
One can observe how secondary driving significantly influenced the number of quantized steps. In Figs.~\ref{fig:2}(a) and \ref{fig:2}(b), we also show the results of the single-mode Floquet QME using {\color{black} dot-dashed} cane-colored lines for comparison. This is done by setting $A_1\!=\!0.5$ eV and $A_2\!=\!0$ eV and can be considered as one of the simplest validity checks for the presented two-mode Floquet QME as it reproduces {\color{black} the results of previous} work~\cite{mosallanejad2024floquet}.
In Figs.~\ref{fig:2}(c) and \ref{fig:2}(d), we plot the time-dependent charge number for the two frequencies, corresponding to selected steps from Fig.~\ref{fig:2}(a). Similarly, in Figs.~\ref{fig:2}(e) and \ref{fig:2}(f), we plot the time-dependent left-current for two chosen $\omega_2$, corresponding to the selected steps from Fig.~\ref{fig:2}(b). Here, we highlight that the oscillatory behavior of $\langle \hat{n}_S \rangle(t)$ requires setting $\mathscr{T}\!=\!4T_1$ and $\mathscr{T}\!=\!8T_1$ for $\omega_2 \!=\!0.750$ eV and $\omega_2 \!=\!0.875$ eV, respectively. The vertical dashed lines show the time intervals $4T_1$ and $8T_1$ [$8T_1$ and $16T_1$] in Figs.~\ref{fig:2}(c) and \ref{fig:2}(e) [Figs.~\ref{fig:2}(d) and \ref{fig:2}(f)]. The integer ratios between the two frequencies are $4/3$ and $8/7$. This indicates that the period by which the observables oscillate will be determined by the in-resonance frequency, $\omega_1$. Compared to the single-frequency case (dot-dashed lines), two key differences in the results can be reported. Firstly, the number of conductance steps increased noticeably, so that for the case of $\omega_2 \!=\!0.750$ eV, there are six large steps and four small steps. For the case of $\omega_2 \!=\!0.875$ eV, there are 14 steps in total. Secondly, observables are no longer static variables but rather quasi-steady.
\begin{figure}[t]
	\begin{center}	\includegraphics[width=7.2cm]{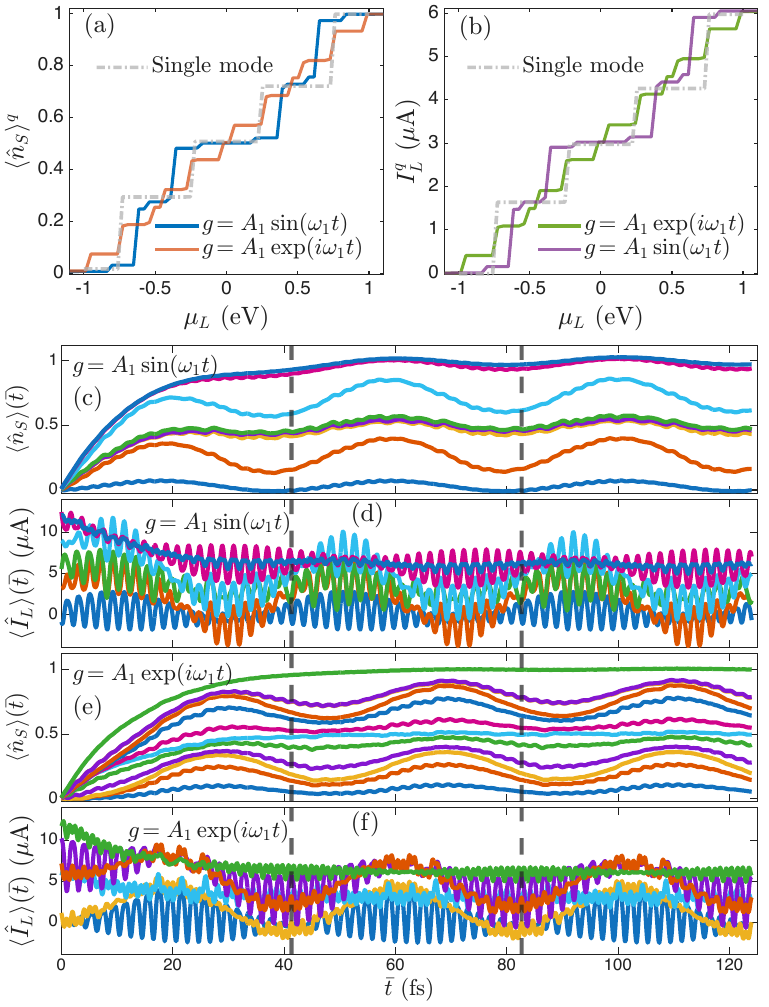}
	\end{center}
    \caption{\label{fig:3} (color online). {\color{black}  Quasi-steady state and dynamics of an open two-level quantum dot driven by multiplying the main resonant driving by an envelope function $\sin^2(\omega_2 t)$. (a) Quasi-steady charge number for primary drivings $g\!=\!A_1\sin(\omega_1 t)$ and $g\!=\!A_1\exp(i\omega_1 t)$. (b) Quasi-steady left current for the two $g$. (c) Time-dependent charge number for the real-valued driving at few plateaus in (a) and (b). (d) Same as (c) for time-dependent left current. (e,f) Analogous to (c,d) for the complex-valued driving. Curves of identical color in (c) [(e)] and (d)[(f)] correspond pairwise.} }
\end{figure}

 \emph{2. Multiplicative term scenario:}
In the second set of examples, we consider multiplying a secondary far-off-resonance driving term as $q(\omega_2,t)\!=\!\sin^2(\omega_2 t)$, to the main in-resonance driving, $g(\omega_1,t)$, i.e., $\odot\!=\!\times$. We then take two forms for $g(\omega_1,t)$ as: (I) the real-valued $g\!=\!A_1\sin(\omega_1 t)$ and (II) the complex-valued $g\!=\!A_1\exp(i\omega_1 t)$. The main amplitude and frequency, ($A_1,\omega_1$), remains unchanged while the second frequency is fixed at $\omega_2\!=\!0.1$ (ten times smaller, far off-resonance) ensuring that $q(\omega_2,t)$ acts as an envelope function. These two driving scenarios can be attributed to the laser sources ($A_1\sin(\omega_1 t)$ [$A_1\exp(i\omega_1 t)$] for linearly [circularly] polarized) that are modulated by passing them through an electro-optic modulator driven at a slower frequency $\omega_2$. In Figs.~\ref{fig:3}(a) and \ref{fig:3}(b), we present the two main quasi-steady observables for the real and complex-valued forms of the primary driving function, $g$. Clearly, the complex-valued driving increased the number of plateaus in an intricate way. In Figs.~\ref{fig:3}(c) and \ref{fig:3}(e) [\ref{fig:3}(d) and \ref{fig:3}(f)], {\color{black}  we plot} the time-dependent charge number [left-current] associated with these two driving scenarios for a few selected steps. Here, the oscillation behavior of the dynamical observables requires to set $\mathscr{T}\!=\!T_2$ as we indicated the sampling time by the last vertical dashed-lines in Figs.~\ref{fig:3}(c)-\ref{fig:3}(f). These dynamical results are interesting, particularly Figs.~\ref{fig:3}(c) and \ref{fig:3}(e), because it shows how the rise and fall of the slow envelope function influence the semi-steady behavior of the time-dependent charge number. We remind the reader that the time-dependent charge number shows an almost steady behavior when the dot is driven only by a single-frequency term~\cite{mosallanejad2024floquet}. Comparison of single-mode and double-mode results demonstrates how incorporating an envelope function in the driving modifies the quasi-steady characteristics.      
\subsection{\label{sec3_subsec:2} Transport through a {\color{black}  driven two-level dot with spin degrees of freedom}}
\begin{figure}[t]
	\begin{center}	\includegraphics[width=7.2cm]{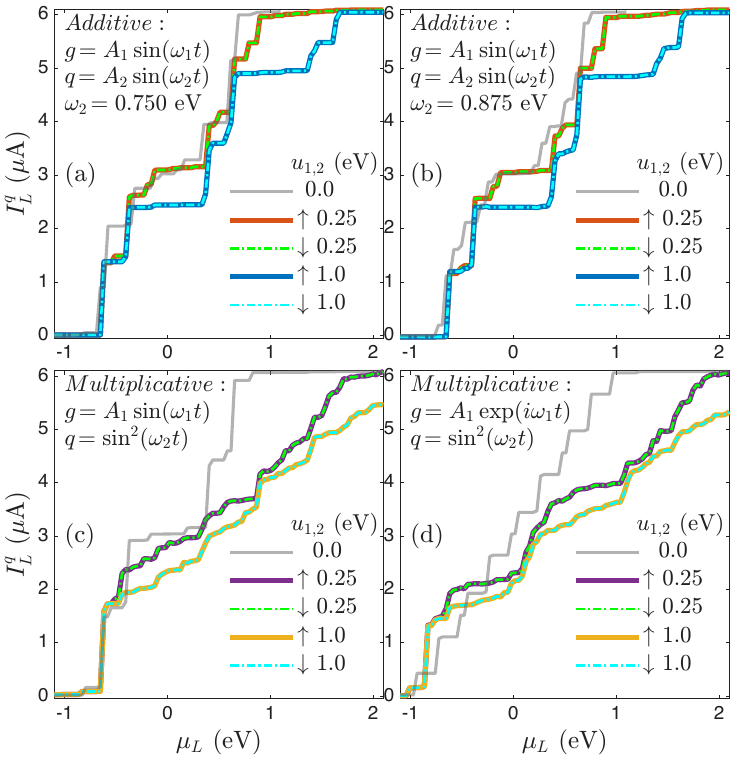}
	\end{center}
\caption{\label{fig:5} (color online). (a) Quasi-steady spin-resolved left-current for the additive term two-mode driving scenario with $\omega_2=0.750~\mathrm{eV}$. (b) Same as (a) for $\omega_2=0.875~\mathrm{eV}$. Here, the solid gray line corresponds to the absence of the e-e interaction. (c) Quasi-steady left-current for the multiplicative scenario when $g\!=\!A_1\sin(\omega_1t)$. (d) Same as (c) when $g\!=\!A_1\exp(i\omega_1t)$.}
\end{figure}

{\color{black}  Here, we extend all previous driven spinless examples to include spin degrees of freedom}. This, for example, requires setting $h_{12/21,\uparrow/\downarrow}(t) \!=\!A_1\sin(\omega_1 t)\!\times\!\sin^2(\omega_2 t)$, and $h_{11/22,\uparrow/\downarrow}\!=\!\epsilon_{1/2}$ for the real-valued multiplicative scenario. Importantly, we will also add the Coulomb electron-electron (e-e) interaction term as $\hat{H}^{ee}\!=\!\sum_{i} u_{i} \hat{d}_{i,\uparrow}^\dagger \hat{d}_{i,\uparrow} \hat{d}_{i,\downarrow}^\dagger \hat{d}_{i,\downarrow}$ to the system's Hamiltonian in Eq.~(\ref{eq:27}). $u_{i}$ describes the energetic cost of adding a secondary electron with a spin opposite to the level $i$. The numerical calculation follows by considering spin-up and spin-down dissipators in the evolution of the density matrix and evaluating spin-resolved observables. All parameters remained unchanged. However, we have used two values for $u_{i}$. Quasi-steady left-current results are presented in Figs.~\ref{fig:5}(a)-\ref{fig:5}(d). In the limit of $u_{i}\!=\!0$, the spin-resolved results are simply identical to the results given previously. 
The main consequence of e-e interactions in quasi-steady observables is to reduce the height of conductance steps. In general, the interplay between drivings and e-e interactions does not induce spin polarization in the current. Interestingly, in the additive cases [fig.~\ref{fig:5}(a) and \ref{fig:5}(b)], quantized steps are recovered in the strong e-e interaction limit ($u_{1,2} = 1.0~\mathrm{eV}$). In contrast, for multiplicative cases [Fig.~\ref{fig:5}(c) and \ref{fig:5}(d)], stronger e-e interactions destroy the driving-induced mini steps. Additionally, in the complex-valued multiplicative case [Fig.~\ref{fig:5}(d)], e-e interactions moderately enhance current flow in low-bias regimes.  
\section{Conclusion} 
\label{sec:4}
In summary, we developed a general two-mode Floquet QME framework capable of accounting for a class of complex driving scenarios, where two oscillating terms jointly drive a multi-level open quantum system. In particular, we have explicitly derived the two-mode Floquet Liouville-von Neumann equation for the Floquet density operator which is exact. Next, we establish a connection between this equation and the two-mode Shirley's-like time evolution operator in Hilbert space. This time evolution formula is a crucial tool that enables us to derive a series of time-local dissipators for a multi-level open quantum system that is driven simultaneously by two independent oscillating terms. Unlike Lindblad-based QMEs, our method avoids the rotating-wave approximation, preventing information loss. However, it possibly cannot accurately capture non-Markovian effects under strong system-bath coupling or extreme low temperatures. The complete validation will be deferred to future work. Additionally, we have made a clear connection between the equation of motion and the time-averaged observables, and demonstrated the application of our two-mode Floquet QME through a number of quantum transport examples. Compared to our previous work on spinless single-mode driven open quantum system, we have found that including a secondary oscillatory driving gives rise to increasing the number of quantized plateaus such that one may engineer the appearance of conductance steps by choosing appropriate parameters within different two-mode driving scenarios. In addition, we show that a certain two-mode complex-valued driving scenario results in a significant increase in quantized plateaus. Furthermore, we have {\color{black} fairly} explored how different driving scenarios affect the dynamical behavior of the observables. {\color{black}  Taking into account the spin degrees of freedom, we show that a moderate electron–electron interaction} mainly results in lowering the hight of few quantized plateaus. We expected that the newly developed two-mode Floquet QME can be applicable in a large number of dissipative systems, as long as the system-bath coupling is not so strong since our two-mode Floquet QME is obtained by well-justified approximations. Although the present work mainly focuses on transport studies at finite temperature and arbitrary voltage, it may also be applicable to areas including quantum thermodynamics, Floquet engineering, quantum information, and strongly correlated quantum systems.

{\color{black} 
\section*{Data Availability}  
The data that support the findings of this study can be available upon reasonable request.}
\begin{acknowledgments}
{\color{black} We thank Lei Sun for very useful conversations.} This work is supported by the National Natural Science Foundation of China (nos. 22361142829 and 22273075), the Zhejiang Provincial Natural Science Foundation (no. XHD24B0301). V.M. acknowledges the funding from the Summer Academy Program for International Young Scientists (Grant No. GZWZ[2022]019).   
\end{acknowledgments}
\bibliography{A0_BiBTMFQMEqttms}
\end{document}